\documentclass{article}

\usepackage[margin=1in]{geometry}
\usepackage[style=authoryear,natbib, backend=biber]{biblatex}
\usepackage{graphicx, amsmath, bm, amssymb, color, xcolor}
\usepackage{cleveref}
\DeclareMathOperator*{\argmax}{arg\,max}

\title{Identifiability, Sensitivity, and Genetic Algorithms in Bacterial Biofilm Selection Models}
\author{Stephen Williams$^1$, 
Daravuth Cheam$^{2,3}$, 
Michele K. Nishiguchi$^2$,\\
Suzanne S. Sindi$^1$, 
Shilpa Khatri$^1$, 
and Erica M. Rutter$^{1*}$\\ 
$^1$University of California, Merced, Department of Applied Mathematics\\
$^2$University of California, Merced, Department of Molecular and Cell Biology\\
$^3$University of California, Merced, Quantitative Systems Biology Graduate Group\\
$^*$Corresponding Author: erutter2@ucmerced.edu}
\date{}

\begin{document}

\maketitle

\begin{center}
\end{center}

\begin{abstract}
Bacteria often develop distinct phenotypes to adapt to environmental stress.
In particular, they can produce biofilms, dense communities of bacteria that live in a complex extracellular matrix. 
Bacterial biofilms provide a safe haven from environmental conditions by distributing metabolic workload and allowing them to perform complex multicellular processes. 
While previous studies have investigated how bacterial biofilms are regulated under laboratory conditions, they have not considered (1) the data requirements necessary to estimate model parameters and (2) how bacteria respond to recurring stressors in their natural habitats.
To address (1), we adapted a mechanistic population model to explore the dynamics of biofilm formation in the presence of predator stress, using synthetic data. 
We used a Maximum Likelihood Estimation framework to measure crucial parameters underpinning the biofilm formation dynamics. 
We used genetic algorithms to propose an optimal data collection schedule that minimised parameter identifiability confidence interval widths.
Our sensitivity analysis revealed that we could simplify the binding dynamics and eliminate biofilm detachment.
To address (2), we proposed a structured version of our model to capture the long-term behaviour and evolutionary selection. 
In our extended model, the subpopulations feature different maximal rates of biofilm formation. 
We compared the selection under different predator types and amounts and identified key parameters that affected the speed of selection via sensitivity analysis.
\end{abstract}

\section{Introduction}

Many bacteria are cosmopolitan, but can survive under extreme conditions such as high or low temperature, humidity, salinity, and pH \parencite{wang_survival_2015,merino_living_2019,kochhar_perspectives_2022}.
Bacteria have evolved to employ several key strategies to maintain the flexibility needed to address such various environmental challenges \parencite{coker_all_2023,marzban_extremophiles_2025,somayaji_insight_2022}.
One such strategy is to adopt distinct ecotypes, observable differences between members of the same species in response to their environment.
One type of adaptive response is the ability to form biofilms in response to both abiotic and biotic conditions.

Biofilms are communities of bacteria spread throughout an extracellular matrix, composed of secreted substances, particularly polysaccharides, proteins, and DNA \parencite{flemming_biofilms_2016,flemming_biofilm_2010}. 
Biofilm formation can confer various forms of resilience to its occupants by creating spatially favourable distributions.
This enables the bacteria to collectively share metabolic workload, and benefit from cooperative behaviours like quorum sensing \parencite{zhang_spatially_2024,li_quorum_2012}.

Free-living or planktonic bacteria move via diffusion or active swimming and readily adhere to surfaces or each other. 
These adhered bacteria become centres for more bacteria to attach, forming microcolonies that grow into mature biofilms through further attachment and internal replication.
Researchers have examined the formation process using experiments \parencite{merritt_growing_2006}, numerical simulations using hydrodynamic theory \parencite{zhao_three-dimensional_2017}, ordinary and partial differential equation models\parencite{trubenova_modeling_2022,dacunto_free_2021}, and agent-based modelling \parencite{nagarajan_agent-based_2022}. 
These studies can provide valuable insights for several applications, including the prevention of biofouling \parencite{elumalai_biofilm_2024,liu_chemical_2024, flemming_biofouling_2020}, infections in medical devices \parencite{hoiby_antibiotic_2010}, and applications to bioremediation \parencite{muhammad_beyond_2020}. 
By understanding biofilm formation dynamics under environmental stress, we also gain a crucial understanding of the wider ecological role of bacteria \parencite{cohen_adaptation_2019,cohen_adaptation_2020,hall-stoodley_bacterial_2004}. 

Differential equation models, where dynamics reflect expected \textit{in vivo} behaviours, offer one way to model biofilm formation dynamics \parencite{bottomley_pde-ode_2024,Garde2020Differential}.
In our study, we use these models to gain a deeper understanding of biofilm formation, particularly under external stressors. 
Our study addresses the types and amounts of data required to parameterise such models, as well as how existing forms of models can be adapted to capture long-term effects that are maintained through selection.
Initially, we will focus on the model presented in \parencite{seiler_grazing_2017}, which utilises a set of differential equations to model a biofilm formation experiment under predation.
The mathematical dynamics governing each population within such models follow established behaviours \parencite{Murray2002}.
While these models are intuitively appealing due to their mechanistic interactions, they also make strong assumptions and introduce numerous potentially unknown parameters \parencite{gunawardena_models_2014,villaverde_assessment_2023}.
As such, understanding how to measure these parameters and quantify their uncertainties is imperative \parencite{kirk_model_2013,simpson_modelling_2024}.
Therefore, we will determine a practical experimental setup that optimises the measurability of the parameters underlying our model.
Additionally, this study enables us to explore how varying external stressors for such systems can lead to adaptations within the population through selection.
The methodologies applied in this study allow us to develop a long-term model from a short-term model to study selection dynamics.

Our first aim is to evaluate this model using structural identifiability, practical identifiability, and sensitivity analysis, and use that information to propose an optimal data collection schedule \parencite{banks_modeling_2014}.
Assuming that the model parameters can be categorised into \textit{a priori} known and unknown categories, we determine the values and associated uncertainties of the unknown parameters.
We validate the inverse problem on synthetic datasets generated from a known baseline parameter set, a common approach in such studies \parencite{burger_computational_2023,liu_parameter_2024}.
The resulting re-estimated parameters and their confidence intervals are calculated from the synthetic dataset.
Using relative practical identifiability, we use a genetic algorithm to propose optimal data sampling schedules.
Further, we employ sensitivity analysis to determine which parameters drove significant differences in population dynamics.
Finally, informed by the sensitivity and identifiability analysis results, we propose a simpler model that produces similar dynamics.

Our second aim is to model bacterial adaptation in a controlled setting. 
Researchers often use long-term evolutionary experiments (LTEE) to study these phenomena \parencite{lenski_experimental_2017,lenski_revisiting_2023}. 
LTEE involves the culture of numerous, initially identical, bacterial populations. 
Periodically subcultured populations in the LTEE allow the bacteria to proliferate continuously, resulting in many generations.
By exposing subsets of these isolated cultures to fixed stressors, their development is guided by those selection pressures.
The bacteria, in turn, will evolve different dynamics to offset the stressors.
In a differential equation model, adaptation across generations can be thought of as bacteria changing the values of their fixed model parameters, or through generic functions that interpolate between conditions \parencite{lewkiewicz_temperature_2022}. 
Assuming our model's inverse problem is well-posed, experimental data could be used to perform the inverse problem at different generations of the LTEE, and then measure the various parameter values.
However, how these parameter values change in this scenario is spontaneous, and there is no measurement of how they transition from their initial to final values.
Since the initial model lacks a mechanism that enables these long-term changes to occur continuously, we will introduce a structured variant of this model, where the bacterial community is comprised of individuals with intrinsic differences in their dynamics. 
We will demonstrate using a systematic resetting of the population, mirroring an LTEE, that selection is then possible for such a structured model.

\Cref{Sect:Models} introduced the two models used throughout the study.
In \Cref{Sect:Model1} a short-term model is outlined, and in \Cref{Sect:Model2} we outline a modified long-term model.
\Cref{Sect:Methods} introduced all the mathematical techniques applied throughout the results section. 
This section introduces structural and practical identifiability, the genetic algorithms techniques, and the sensitivity analysis.
\Cref{Sect:Results} discusses our results. 
We first outline the results of the short-term model's structural and practical identifiability, before proposing a genetic algorithm approach to improve the model's identifiability. 
Next, we introduce the long-term model in \Cref{res:struct}, and detail the sensitivity analysis of this model in \Cref{res:sena2}.
Finally, a discussion of the results is given in \Cref{sec:conc}.

\section{Population Models}
\label{Sect:Models}
\subsection{Short-term Biofilm Formation Model} \label{Sect:Model1}

\begin{figure}[ht!]
\centering
    \includegraphics[width=\linewidth]{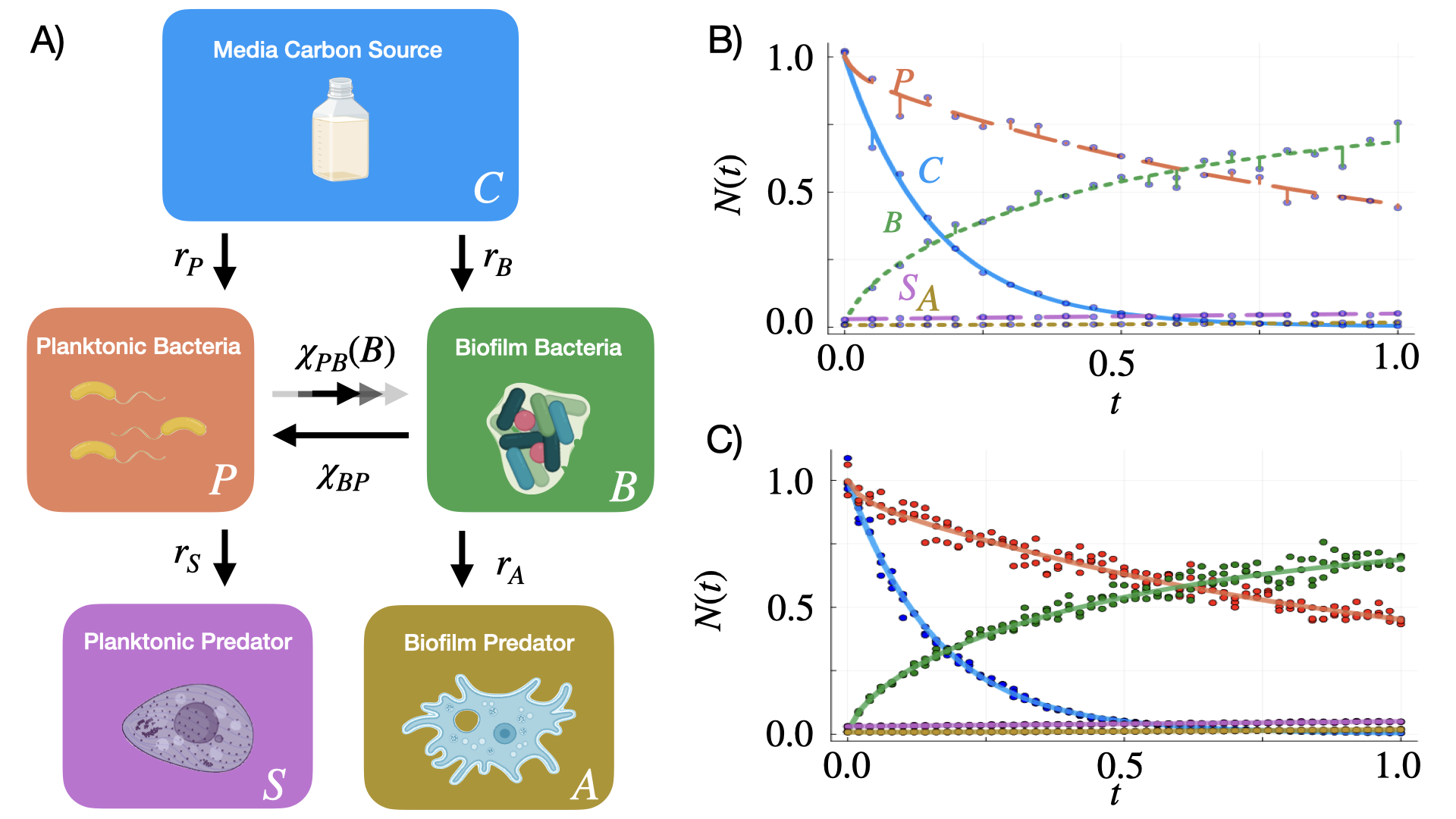}
    \caption{\textbf{Model schematics and example solutions.}
    (A) Schematic of the short-term ODE system (\Cref{eqn:ODE_C}-\Cref{eqn:ODE_A}). 
    Both bacterial phenotypes consume the media carbon source, $C$, to reproduce, with growth rates $r_P$ for the planktonic cells $P$ and $r_B$ for the biofilm cells $B$.
    Bacteria transition from a planktonic to a biofilm phenotype at a rate dependent on biofilm size, $\chi_{PB}(B)$.
    The biofilm bacteria detach, becoming planktonic at a constant rate $\chi_{BP}$.
    The bacteria are consumed by their associated predators.
    (B) An example solution (solid, dashed, and dotted lines) with initial conditions  $[C_0,P_0,B_0,S_0,A_0] = [1.0,1.0,0.01,0.03,0.008]$ and an example synthetic dataset  (circular purple points). 
    Here, $N(t)$ is the population size of each compartment in the model at time $t$.
    Vertical lines indicate the distance between the synthetic data and the noise-free solutions.
    (C) The best-fit solution (solid lines) compared to moderately noisy synthetic data (scatter points) for the system in (B). 
    $N(t)$ again shows the population size of each compartment.
    The synthetic dataset consists of three biological replicates, each with 51 uniformly spaced time points.
    The fit line (dashed lines) is the resulting output from the estimated values of $[\chi_{PB}^{\max}, \chi_{PB}^{\min}, a, \chi_{BP}]$.
    }
    \label{fig:figure1}
\end{figure}

In the first part of our study, we focus on the dynamics of a population within one self-contained experimental setting.
To represent the flow of nutrients through our biofilm-forming system, we use a stoichiometric mathematical model proposed in \parencite{seiler_grazing_2017}, see Figure \ref{fig:figure1}A.
The different populations are measured using state variables, which indicate the carbon content contained in these populations, access to which is limited by the initial content present \parencite{bren_last_2013,hornung_quantitative_2018}. 
Our equations, which model the state variables, are given by 

\begin{align}
    \frac{dC}{dt} &= 
    - \underbrace{r_P \frac{C}{C+H_C} P}_{\substack{\text{Carbon consumed}\\ \text{by Planktonic}}}
    - \underbrace{r_B \frac{C}{C+H_C} B}_{\substack{\text{Carbon consumed}\\ \text{by Biofilm}}}, \label{eqn:ODE_C}\\
    \frac{dP}{dt} &= 
    \underbrace{e_b r_P \frac{C}{C+H_C} P}_{\substack{\text{Planktonic growth}}}
    - \underbrace{r_S\frac{P}{P+H_P}S}_{\substack{\text{Planktonic consumed}\\ \text{by predator}}}
    - \underbrace{\frac{a\chi_{PB}^{\max} + B\chi_{PB}^{\min}}{a + B}P}_{\substack{\text{Planktonic attaching}\\ \text{to Biofilm}}} 
    + \underbrace{\chi_{BP}B}_{\substack{\text{Biofilm}\\ \text{Detachment}}} , \label{eqn:ODE_P}\\
    \frac{dB}{dt} &= 
    \underbrace{e_b r_B \frac{C}{C+H_C} B}_{\substack{\text{Biofilm growth}}} 
    - \hspace{1.5mm} \underbrace{r_A\frac{B}{B+H_B}A}_{\substack{\text{Biofilm consumed}\\ \text{by predator}}}
    \hspace{2mm}+ \underbrace{\frac{a\chi_{PB}^{\max} + B\chi_{PB}^{\min}}{a + B}P}_{\substack{\text{Planktonic attaching}\\ \text{to Biofilm}}} 
    - \underbrace{\chi_{BP}B}_{\substack{\text{Biofilm}\\ \text{Detachment}}} , \label{eqn:ODE_B}\\
    \frac{dS}{dt} &= 
    \underbrace{e_S r_S\frac{P}{P+H_S}S}_{\substack{\text{Planktonic predator}\\ \text{growth}}} , \label{eqn:ODE_S}\\
    \frac{dA}{dt} &= 
    \underbrace{e_A r_A\frac{B}{B+H_A}A}_{\substack{\text{Biofilm predator}\\ \text{growth}}}. \label{eqn:ODE_A}
\end{align}
Here, the state variables represent the compartments present in the system: the media source ($C$), planktonic bacteria ($P$), biofilm bacteria ($B$), planktonic predator ($S$), and biofilm predator ($A$).
The carbon source (\Cref{eqn:ODE_C}) is consumed by both planktonic and biofilm cells. 
\Cref{eqn:ODE_P} and \Cref{eqn:ODE_B} represent  planktonic and biofilm bacteria, respectively. 
Their similar structure includes (a) carbon-limited growth, (b) predation by their respective predators, (c) biofilm attachment, and (d) biofilm detachment. 
The attachment of planktonic bacteria to biofilm depends on the size of the biofilm: the bacteria attach at a maximum rate $\chi_{PB}^{\max}$ for small biofilm population sizes, but when biofilms are large, the rate reduces to $\chi_{PB}^{\min}$. 
The parameter $a$ determines the biofilm size at which the attachment rate transitions between these values.
The detachment rate is proportional to the size of the biofilm, with a constant rate $\chi_{BP}$. 
\Cref{eqn:ODE_S} and \Cref{eqn:ODE_A} represent the planktonic and biofilm predators, which grow from consuming their respective food sources. 
In \Cref{eqn:ODE_C}-\Cref{eqn:ODE_A}, all growth and consumption functions follow a type-II functional response \parencite{gesztelyi_hill_2012}.
A schematic of nutrient flow is given in \Cref{fig:figure1}A.

\Cref{tab:Modelparameters} presents the nominal values of the parameters used throughout our results. 
These values were calculated by non-dimensionalising the values from \parencite{seiler_grazing_2017} (see  \Cref{SubSect:NDprocess}).
The non-dimensionalisation was chosen to provide consistency when comparing the relative sizes of the populations, regardless of whether they are in liquid (media, planktonic, or planktonic predator) or on the surface (biofilm, biofilm predator).
However, our non-dimensional parameters implicitly depend on the geometry of the container used in experiments; thus, they must be re-determined for other experimental setups.
Throughout, when referencing the values in \Cref{tab:Modelparameters}, we will use a tilde (e.g., the nominal value of $\chi_{BP}$ will be denoted as $\tilde\chi_{BP}$). 
Our primary focus is to use this model to investigate changes in biofilm formation.
Thus, we fix all parameters in the model except those most relevant to biofilm formation: $\chi_{\max}, \chi_{\min}, a, $ and $\chi_{BP}$. 
For a given synthetic dataset, our analysis will focus on re-estimating the values of these parameters and determining their uncertainties.

\subsection{Long-term Structured Biofilm Model} \label{Sect:Model2}

The parameters in the model of \Cref{Sect:Model1} are fixed constants, which means the bacteria cannot adopt different dynamics or adapt. 
To study the long-term evolutionary dynamics of bacterial phenotypes, we propose a parameter-structure population model (see the schematic in \Cref{fig:figure6}A).
In this updated model, we assume that instead of having a single population of bacteria (and thus a single value of the constant parameter describing the binding dynamics), we have a collection of similar but slightly different subpopulations.
Within these populations, the value of their maximum binding rate, referred to in the short-term model as $\chi_{PB}^{\max}$, takes on a continuum of values.
The modified set of equations governing the system is:
\begin{align}
    \frac{dC}{dt} &= - \left( r_P \frac{C}{C + H_C}\mathcal{P} + r_B\frac{C}{C + H_C}\mathcal{B} \right), \label{eqn:StrucC}\\
    \frac{dP_i}{dt} &= e_b r_P \frac{C}{C + H_C} P_i - r_S \frac{\mathcal{P}}{\mathcal{P} + H_P}\frac{P_i}{\mathcal{P}}S - \frac{a}{a+\mathcal{B}} \chi_i P_i, ~~~ i\in (1,2, ..., N_c), \label{eqn:StrucP}\\
    \frac{dB_i}{dt} &= e_b r_B \frac{C}{C + H_C} B_i - r_A \frac{\mathcal{B}}{\mathcal{B} + H_B}\frac{B_i}{\mathcal{B}}A + \frac{a}{a+\mathcal{B}} \chi_i P_i, ~~~ i\in (1,2, ..., N_c), \label{eqn:StrucB}\\
    \frac{dS}{dt} &= e_S r_S \frac{\mathcal{P}}{\mathcal{P} + H_P}S, \label{eqn:StrucS}\\
    \frac{dA}{dt} &= e_A r_A \frac{\mathcal{B}}{\mathcal{B} + H_B}A, \label{eqn:StrucA}\\
    \mathcal{P} &= \sum_i P_i, \label{eqn:StrucTotP}\\
    \mathcal{B} &= \sum_i B_i. \label{eqn:StrucTotB}
\end{align}
This set of equations is structurally similar to Equations \eqref{eqn:ODE_C}-\eqref{eqn:ODE_A}, with the key difference being that we now have a compartment structure in $P$ and $B$. 
Equations \eqref{eqn:StrucP} and \eqref{eqn:StrucB} describe the dynamics of the bacteria occupying the $i$th compartment. 
The total amount of planktonic and biofilm populations is the sum of the individual compartments (\Cref{eqn:StrucTotP} and \Cref{eqn:StrucTotB}). 
As before, there is growth from media source consumption and elimination due to predation. 
The growth and death terms are scaled to be proportional to the size of the compartment. 
In this updated model, there is no biofilm detachment rate. 
The remaining term concerns the biofilm attachment rate, which remains dependent on the size of the underlying biofilm. 
However, the minimum attachment rate is now zero for large biofilms. 
Each of the $N_C$ ordered compartments has an associated biofilm maximum attachment rate, $\chi_i$, at a uniformly spaced set of values (i.e., $\chi_1 < \chi_2 = \chi_1 + \Delta\chi < ... < \chi_N = \chi_1 (N-1)\Delta\chi $). 
The bacteria are assumed to retain their maximum attachment rate $\chi_i$ upon transitioning between the phenotypes; thus, $P_i$ and $B_i$ are coupled, but do not interact with the other compartments.
For clarity, we use the notation $\chi_{\max}$ in figures and discussions to represent the maximum value of the compartment. 
We select our range of $\chi_i$ so that evolving populations never reach the endpoints $\chi_1$ or $\chi_N$ \parencite{rutter_continuous_2017}.
To represent a continuum of $\chi_{\max}$ values, we used $N_c=2000$ compartments in our results.
The dynamics for the carbon media source (\Cref{eqn:StrucC}) and both predators (\Cref{eqn:StrucS} and \Cref{eqn:StrucA}) are analogous to the previous model. 

To solve Equations \eqref{eqn:StrucC}-\eqref{eqn:StrucTotB}, the initial conditions are the same for $C$, $S$, and $A$, but the initial condition for $P_i$ or $B_i$ must be a distribution.
To solve our system, we take the initial distribution:

\begin{equation}
    P_i = \frac{A}{\sigma_\chi} \exp\left(-\frac{\left(i-i_c\right)^2}{2\sigma_\chi^2} \right),
\end{equation}
where $A$ is a constant chosen so that the sum across all compartments $\sum_i^{N_c}P_i=1$.
The distribution has some width, measured by the standard deviation $\sigma_\chi$, and a mean $i_c$ corresponding to a binding rate of $\chi_{i_c}$.
For convenience, we use $B_i = 0.001 P_i$ as the initial condition for the biofilm compartment to avoid issues related to calculations involving the total biofilm size. 

To mimic the LTEE setup, the structured model is repeatedly solved to $t=1$, representing one generation, and re-seeded. 
This choice is motivated by an envisioned LTEE experimental setup in which the supernatant (containing media, planktonic predators, and biofilm predators) is removed from the system. 
Following this, the preserved biofilm is mechanically broken up to become the new planktonic population (e.g., initial conditions $P_i = B_i/\sum_i B_i$ and $B_i = 0.001 P_i$). 
New media and an appropriate level of predators are then reintroduced (e.g., $C$, $S$, and $A$ are reset to their initial values). 
By iterating this process, simulating the system to $t=1$ and then resetting the system using only the biofilm population, selection is enabled across multiple generations.

\section{Methods} \label{Sect:Methods}

The dynamics of biofilm formation under predation are explored using the above population models.
In this section, we outline the details of the techniques used to simulate these dynamics and describe the tools used to analyse model outputs.

\subsection{Model Numerical Solutions and Synthetic Data Construction}
\label{SubSect:SynthData}
The mathematical models of the populations (see \Cref{Sect:Model1} and \Cref{Sect:Model2}) are solved numerically using the Julia package DifferentialEquations \parencite{rackauckas2017differentialequations}, specifically the Tsit5 implementation of the Runge-Kutta 4(5) \parencite{tsitouras_rungekutta_2011}. 

We utilise noisy synthetic datasets to assess our ability to measure model parameters at various levels of sparsity.
To generate our synthetic data for compartment $i$, $Y_i(t)$, we use the following statistical model:
\begin{equation} \label{eqn:statModel}
    Y_i(t) = X_i(t) \mathcal{E}_i, ~~~~~\text{where} ~~~~~ \mathcal{E}_i \sim \text{Log-Normal}(\mu,\sigma).
\end{equation}
For each data point $X_i(t)$, we modify the exact numerical solution according to multiplicative noise drawn from a Log-normal distribution \parencite{murphy_implementing_2024}. 
A synthetic dataset example is overlaid on the numerical solution in \Cref{fig:figure1}B. 
We select the log-normal distribution to preserve the positivity of the synthetic data. 
The value of our Log-normal is given by $\mathcal{E}_i = \exp(\mu_l + \sigma_l \epsilon_i)$ where $\mu_l$ and $\sigma_l$ are some real numbers and $\epsilon_i$ is a standard normally distributed random variable (i.e., mean zero and standard deviation one).
These values $\mu_l$ and $\sigma_l$ tell us the mean and standard deviation of $\log(\mathcal{E}_i)$, not of $\mathcal{E}_i$.
We represent our distribution's mean and standard deviation as $\mu$ and $\sigma$, respectively.
The relationship between these values is as follows, 
\begin{align}
    \mu_l = \log\left(\frac{\mu^2}{\mu^2 + \sigma^2}\right), \label{eqn:LN_mean}\\
    \sigma_l^2 = \log\left(1 + \frac{\sigma^2}{\mu^2}\right). \label{eqn:LN_std}
\end{align}
Using the Julia packages Distributions and Random, the values $\mu_l$ and $\sigma_l$ were used to generate samples from the Log-normal distributions \parencite{JSSv098i16}.
We assume that such multiplicative noise is appropriate to capture the errors present in measurements of each compartment, and that the values of $\mu$ and $\sigma$ are the same for each compartment.
While we assume we know the value of $\sigma$ \textit{a priori}, it can easily be estimated by extending our parameter estimation approach if it's not known in advance \parencite{simpson_practical_nodate}.

In experiments, the measurement of bacterial population size through optical density or Colony-Forming Units is often a balance between speed and accuracy.
Modern techniques report coefficients of variation in the range of 1-5\% (corresponding to values as low as $\sigma=1\times10^{-2}$) \parencite{rahman_counter--chip_2024,martini_maximum_2024}. 
In \Cref{res:ident}, we use an artificial value of $\sigma = 2.5 \times 10^{-3}$ to compare the relative difference between parameter estimation bounds, particularly in cases with very limited datasets.
This noise represents an absolute best-case scenario value for these parameter estimate confidence bounds.
At this noise level, we observe a range of practical identifiability across our parameters of interest, with some parameters estimated with very narrow 95\% confidence intervals, while others have wide intervals.

\subsection{Model Structural Identifiability} \label{SubSect:SIdent}

The first step in model evaluation is determining if the system is structurally identifiable.
This intrinsic model property determines whether, given its noise-free outputs, the parameters within the model can be re-determined uniquely \parencite{villaverde_assessment_2023}.
To determine whether the models presented are structurally identifiable, we used the online Maplecloud Structural Identifiability Toolbox \parencite{Hong2020Global}.
This program uses the Structural Identifiability Analyser (SIAN) \parencite{hong_sian_2019}.
We determined whether model parameters were globally identifiable by providing different information availability.
For example, in real-world experiments, the concentration of planktonic bacteria, the concentration in the biofilm, and the number of predators are relatively straightforward to measure (using Colony Forming Unit measurements (CFU), optical density (OD), haemocytometry, etc.) \parencite{pan_comparison_2014, beal_robust_2020, zhang_improvement_2020}.
However, measuring the concentration of nutrients in the media, which can be done using techniques like liquid chromatography mass spectrometry, is comparatively impractical or destructive \parencite{floris_lcmsms_2019}.
Restricting access to information about the media in a dataset (i.e., omitting synthetic data regarding $C$ throughout simulations) may pose issues regarding what parameters can be determined.
Thus, structural identifiability analysis allows us to determine whether information about parameters within the system is even possible.

\subsection{Model Practical Identifiability} \label{SubSect:PIdent}

Throughout our study, we will assess for given data scenarios whether the model is practically identifiable.
This analysis allows us to determine the parameters within the model and quantify the statistical evidence that these measurements are significant.

We use the Maximum Likelihood Estimator (MLE) for a given set of synthetic data to re-estimate the baseline parameters used to generate the synthetic data \parencite{simpson_parameter_2025}.
To do this, suppose we have data $Y_{i,j}(t_k)$ at a specific time $t_k$, where $i$ represents the state variable and $j$ represents the biological replicate. 
For a given initial guess at the baseline parameters $\bm{\theta}$, we can use our mathematical model to calculate what the parameter guess predicts the state variables should be, $X_i(t_k|\bm{\theta})$.
The probability of the measurement $Y_i(t)$, $p(Y_{i,j}(t_k)/X_i(t_k;\bm{\theta}) ~|~ \text{Log-normal}(\mu,\sigma))$, can be calculated since the distribution of the synthetic data is known.
Working with the log-probability allows us to write the log-probability of the whole dataset as
\begin{equation} \label{eqn:llhood}
    \ell = \sum_{i,j,k} \log[p(Y_{i,j}(t_k)/X_i(t_k;\bm{\theta}))].
\end{equation}
We have dropped the explicit error distribution label inside $p(\cdot)$ for simplicity.
The best parameter set $\bm{\theta}_{opt}$ is the one which maximises the value of $\ell$ (equivalently, it is the parameter set which maximises the probability of measuring the data under our statistical model).
We will refer to the optimal value as $\ell_p$.
Hence,
\begin{equation} \label{eqn:MLEOutput}
    \bm{\theta}_{opt} = \argmax_{\bm\theta} \sum_{i,j,k} \log[p(Y_{i,j}(t_k)/X_i(t_k;\bm{\theta}))].
\end{equation}
To calculate this, we used the NLopt Julia package \parencite{NLopt}.
Using an initial guess of the values of the parameters, the Nelder-Mead algorithm allows the optimal parameter values of \Cref{eqn:MLEOutput} to be found \parencite{galantai_neldermead_2024}. 

Without proper context, these estimated values can be somewhat misleading. 
The model is not practically identifiable with a given dataset if two distinct parameter sets provide the same log likelihood values.
To quantify whether our parameters are practically identifiable, we calculated profile likelihood by fixing one parameter of interest at a time and then re-estimating the remaining parameters \parencite{raue2009structural,simpson_parameter_2025}. 
For example, suppose we have measured the optimal values for our dataset $\bm{\theta}_{opt} = (\bar\theta_1,\bar\theta_2)$.
We can, for some small value $\delta$, construct a new parameter set $\bm{\theta}' = ((1+\delta)\bar\theta_1, \theta_2)$.
Where $(1+\delta)\bar\theta_1$ is fixed, and $\theta_2$ is re-estimated to maximise the log likelihood of our dataset.
In that case, the log likelihood $\ell_p(\delta)$ must necessarily be less than the true optimal $\ell_p$ if $|\delta|>0$.
By examining the value of $\bar\ell_p = \ell_p - \ell_p(\delta)$ across a range of values of $\delta$, we can determine how sensitive the estimation is to univariate variation.
$\bar\ell_p$ is $\chi^2$-distributed \parencite{li_graduate_2019} and, therefore, we can determine a one-degree-of-freedom quantile (for our univariate parameter) above which the probability of the data is equal to or greater than a desired level.
This quantile can be implemented using the Distributions Julia package and has a value $\ell_{crit} = 1.921$ for a 95\% confidence level.
A confidence interval can be constructed by interpolating $\bar\ell_p + \alpha$ and finding its roots.
We refer to this interval as the confidence interval width $CIW_{95\%}$.
The function CubicSplineInterpolation from the Interpolations Julia package, alongside the Roots package's function find\_roots, were used to perform these calculations \parencite{Kittisopikul_Interpolations_jl_2025,Rootsjl}.

\subsection{Genetic Algorithm} \label{SubSect:GenAlgo}

In our log-likelihood parameter estimation in \Cref{SubSect:PIdent}, \Cref{eqn:llhood}, no pre-supposition is made on the sampling times $t_k$.
One obvious schedule choice is uniform; however, different parameters can drive dynamics at different population sizes.
For example, as in \Cref{fig:figure1}B, when there is a tiny biofilm population, the attachment rate $\chi_{PB}^{\max}$ dominates the biofilm population size changes. 
In contrast, $\chi_{PB}^{\min}$ barely has any impact.
To measure $\chi_{PB}^{\max}$ as accurately as possible, it is better to have more data samples when the parameter is more impactful.
We used a genetic algorithm to systematically construct schedules that allow us to measure a given parameter most effectively \parencite{lam_evolving_2024}. 
A flowchart of the algorithm is given in \Cref{fig:figure3}A.

As described in \Cref{SubSect:PIdent}, for a given synthetic dataset, a 95\% confidence interval can be calculated for the best fit parameters.
The genetic algorithm we employed closely follows previous work \parencite{lam_evolving_2024}, but is explained here for clarity.
The genetic algorithm first generates $M=100$ random schedules to determine the best possible sampling schedule.
These schedules consist of a fixed number of points; we generated  11, 21, and 51 sample point schedules in our case.
The sample time points are generated by sampling a uniform distribution on $(0,1]$, with the initial condition always provided at t=0.
Each new sample is initially checked to be greater than $\Delta t=0.01$ (14.4 minutes) away from other sample times.
For each schedule, a synthetic dataset is generated, and the value of $CIW_{95\%}$ is found.
The schedules can then be ranked according to the value of $CIW_{95\%}$, with the smallest $CIW_{95\%}$ values ranked highest.
For the $M$ cases, we use a cloning process to generate a new set of schedules \parencite{lam_evolving_2024}.
The best schedule is preserved unchanged.
The new $i$th schedule in the cloned set is a copy of the $\text{ceil}(i/14)^2$th schedule in the previous set.
This cloning method keeps many copies of the most highly ranked schedules and fewer, less highly ranked ones, up to the 51st best.
Each cloned schedule time point is then modified by a value drawn from a normal distribution with a standard deviation $\sigma = 9\Delta t/N_{epochs}$.
The values were constrained to remain in the range $(\Delta t,1-\Delta t)$.
Because of the relatively large standard deviation of the cloning noise, points could move freely within this interval, exchange positions, and the minimum distance between scheduled points was not enforced from one epoch to the next.
The example from \Cref{fig:figure3}B and \Cref{fig:figure3}C used 11 schedule sample points (and three biological replicates) across five epochs.
In later examples, three epochs were used to generate the optimal schedules, as in our example, relatively rapid settling of values motivated this approach.

\subsection{Short-term Model Metrics and Sensitivity} \label{SubSect:SenA_1}

We perform sensitivity analysis to assess the relationship between variations in biofilm dynamics parameters and their resulting population. 
Since we are interested in the dynamics of biofilm formation, we chose the final biofilm size as our quantity of interest, which we will refer to as $B^*$.
We used Saltelli Sampling to generate a collection of 40,000 parameter sets, each of which the value of the parameters $\chi_{\max},~\chi_{\min},~a,~\chi_{BP}$ are varied over the range 10\% to 190\% of their nominal values (given in \Cref{tab:Modelparameters}) \parencite{saltelli_making_2002}. 
In our results, we scale the parameters by their nominal value to allow for the best comparison between parameters.

We use a variety of metrics to analyse the relationship between the parameters and $B^*$.
To display the general trend between parameters and $B^*$, we present univariate scatter plots (see  \Cref{fig:figure5}A-D) and their best-fit lines calculated via Julia's CurveFit package. 
To measure linear relationships between the parameters and $B^*$, we calculate the Pearson and Partial correlations, the latter of which controls for the variation of other parameters. 
We also measure Spearman and Partial Rank Correlations, which allow us to remove the effects driven by outliers and measure the presence of monotonic relationships across the entire dataset. 
For a more in-depth discussion of these metrics, see \parencite{marino_methodology_2008}. 
The above metrics were calculated using the Julia package GlobalSensitivity \parencite{dixit2022globalsensitivity}.

Finally, we use Sobol Sensitivity indices to decompose the variance in the full dataset \parencite{sobol_global_2001}.
Sobol indices examine the magnitude of the present variance, even if the relationships between parameters are non-linear. 
Additionally, Sobol sensitivity provides a method for examining the collective effects of parameters, assessing potential relationships between them by measuring second-order indices.

\subsection{Long-term Model Metrics and Sensitivity} \label{SubSect:SenA_2}

In the long-term adaptation model, we apply sensitivity analysis methods to examine the relationship between model parameters and the change in biofilm binding rate.
The average binding value can be found using
\begin{equation} \label{eqn:MBR}
    \bar{\chi}_{\max} = \frac{1}{\mathcal{B}} \sum_j\chi_j B_j.
\end{equation}
Applying Equation \eqref{eqn:MBR} to the initial condition for the biofilm, we can get the initial biofilm average binding $\bar{\chi}_{\max}^i$ (which corresponds to $\chi_{i_C}$ from \Cref{Sect:Model2}). 
We then recalculate the average binding rate after 50 generations, $\bar{\chi}_{\max}^f$.
Our quantity of interest is the ratio of these rates $\bar{\chi}_{\max}^f/\bar{\chi}_{\max}^i$, which measures the relative impact of selection in the population.
The parameters examined in the sensitivity analysis are the amount of resetting predators $S_0$ and $A_0$, the initial average binding rate $\chi_{i_c}$, and the variability of binding rates in the initial condition $\sigma_\chi$.

Sobol sensitivity was performed for a Saltelli sample using these methods (see \Cref{fig:figure7}D), with 40,960 samples.
The samples in this analysis were generated using the Python Library SALib sample.sobol's function sample.
The simulations were carried out in parallel using MATLAB to optimise the code runtime, the function ODE45 replacing the DifferentialEquations Tsit5 algorithm \parencite{MATLAB_ode45}.
Since each sample's execution took a relatively similar time, splitting the workload evenly between a MATLAB parallel pool of `threads' workers heuristically resulted in the fastest execution.
We saved the outputs for each parameter set, then analysed them separately using the Python Library SALib \parencite{Iwanaga_Toward_SALib_2_0_2022}.
This library offers convenient functionality for importing parameters and datasets (note that results in Julia's GlobalSensitivity must be generated locally).

\section{Results}
\label{Sect:Results}

Our results fall into two main sections.
In \Crefrange{res:ident}{res:sena}, we will explore the short-term model presented in \Cref{Sect:Model1}.
We aim to determine whether the model parameters can be defined under specific data availability scenarios and assess the uncertainty in these measurements before examining the model's sensitivity to variations in these parameters.
In doing so, we show differences between the identifiability of the parameters and demonstrate that sampling schedule and frequency have a substantial effect on the parameters' confidence intervals.
Next, in \Crefrange{res:struct}{res:sena2}, we propose a new model for our system.
This new model captures the impacts of selection on populations by introducing population structure within the bacterial compartments.
Finally, we will measure this impact under specific conditions through a sensitivity analysis, showing that variability in bacterial parameters and predation type give the most significant changes in the selection dynamics.

\subsection{Measuring the impacts of data sampling frequency on model practical identifiability} \label{res:ident}

Our first aim is to explore the structural and practical identifiability of the short-term model described in \Cref{Sect:Model1}.
To determine the global structural identifiability of our initial mathematical model (\Cref{eqn:ODE_C}-\Cref{eqn:ODE_A}), we used the SIAN toolbox (see \Cref{SubSect:SIdent} for details).
When information on the state variables $[C, P, B, S, A]$ is available, the complete set of parameters and initial conditions present in \Cref{tab:Modelparameters} is globally identifiable.
In general, we will focus on the identifiability of the parameters $(\chi_{PB}^{\max}, \chi_{PB}^{\min}, a, \chi_{BP})$, since these parameters determine the dynamics through which the bacteria transition between planktonic and biofilm phenotypes.
We assume all other parameters are known exactly \textit{a priori}.
When limiting our interest to this subset of parameters, only the biofilm $B$ needs to be measured for global identifiability.

To provide the best-case scenario for measuring these parameters practically, we will assume that we have access to measurements of all state variables.
We generate a synthetic dataset with three biological replicates (described in \Cref{SubSect:SynthData}) based on a uniform sample schedule as a proxy for experimental measurements.
Additionally, the samplings include a measurement of the system's initial condition. 
\Cref{fig:figure1}B displays a synthetic data example using the nominal values from \Cref{tab:Modelparameters} for a single biological replicate with $\sigma=0.05$, comprising 21 equispaced time points (including the initial condition).
Another example with all three biological replicates, the same noise level, and 51 uniformly spaced measurements is presented in \Cref{fig:figure1}C.

\begin{table}[!ht] 
\centering
\begin{tabular}{|c||c|c|c|c|}
\hline
$\sigma$ & $\chi_{PB}^{\max}$ & $\chi_{PB}^{\min}$ & $a$ & $\chi_{BP}$ \\
\hline \hline
Nominal & 3.20  & 0.0320 & 0.0800 & 0.120 \\
\hline
0.0025  & 3.20  & 0.0294  & 0.0803 & 0.117 \\
\hline
0.0050  & 3.59  & 0.2630 & 0.0539 & 0.298 \\
\hline
0.0100  & 3.19  & 0.0212 & 0.0811 & 0.108 \\
\hline
0.0500  & 3.13  & -0.0345 & 0.0869 & 0.055 \\
\hline
\end{tabular}
\caption{
\textbf{Example parameter estimation values at different noise levels} 
For given values of standard deviation measured in the Log-normal output distribution. 
\Cref{SubSect:SynthData} contains the details on the generation of synthetic data and a discussion of error magnitudes. 
Values are estimated using MLE, as described in \Cref{SubSect:PIdent}.
}
\label{tab:fitVals}
\end{table}

We use MLE to estimate the parameters (see \Cref{SubSect:PIdent}) from synthetic data. 
An example of the model with the best-fit parameters is displayed in \Cref{fig:figure1}C.
Qualitatively, we observe a good fit when visually comparing the curves in \Cref{fig:figure1}C to the numerical solution in \Cref{fig:figure1}B.
\Cref{tab:fitVals} compares the recovered parameters for various noise levels up to $\sigma=0.05$. 
Despite visual agreement with the numerical solution, it is clear that the recovered parameter values sometimes differ from the nominal values.
The estimation accuracy improves as the noise level decreases.
For the remainder of this section, we will focus on the very-low noise case, $\sigma=2.5\times10^{-3}$, since it consistently yields a very tight 95\% confidence interval for parameters thought to be practically identifiable. 
This noise level also provides a significant, but measurable, confidence interval at some sampling levels for less identifiable parameters.

\begin{figure}[ht!]
\centering
    \includegraphics[width=\linewidth]{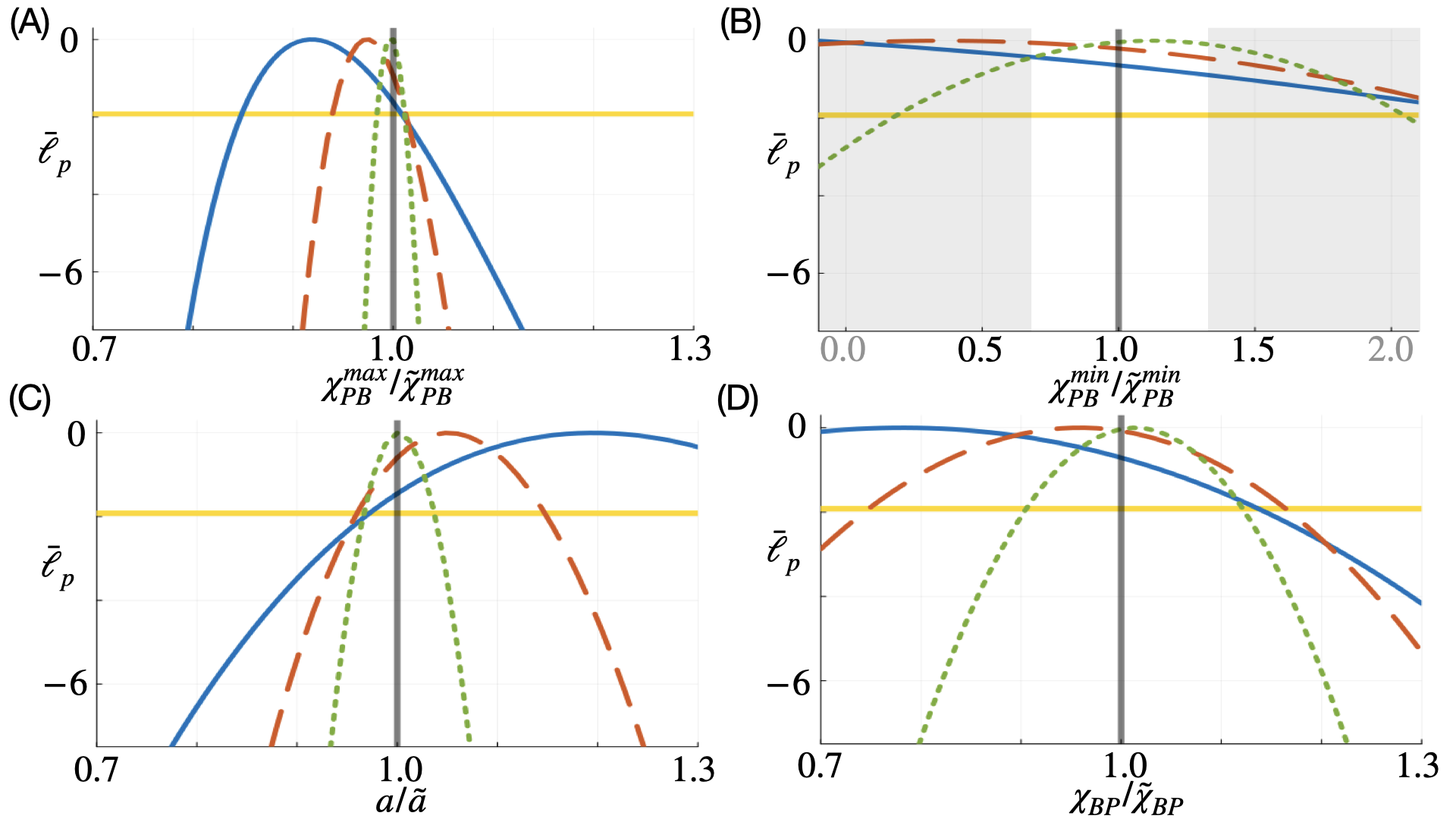}
    \caption{\textbf{Measuring practical identifiability using univariate profile likelihoods.} 
    Log likelihoods relative to optimal solution $\bar{\ell_p}$ of parameters $\chi_{PB}^{max}$ (A),$\chi_{PB}^{min}$ (B), $a$ (C), and $\chi_{BP}$ (D).
    In each case, the $x$-axis is normalised by the parameter's nominal value (see \Cref{tab:Modelparameters})
    and the grey vertical line represents the true value.
    For each parameter, the value at uniformly sampled time points of 11, 21, and 51 is given by the blue (solid), red (dashed), and green (dotted) curves, respectively. Intersections with the yellow curve correspond to 95\% confidence intervals for the parameter values given the data (see \Cref{tab:ConfIntervalWidths} for the confidence interval width values).
    To highlight the difficulty in calculating the confidence interval in (B), vertical lines are given at $\theta/\tilde{\theta} = 0.7$ and $\theta/\tilde{\theta} = 1.3$, the x-axis range over which (A), (C), and (D) are presented.
    }
    \label{fig:figure2}
\end{figure}

We next turn to profile likelihoods to explore parameter identifiability based on equispaced sampling schedules. 
\Cref{fig:figure2} displays the univariate profile likelihood (see \Cref{SubSect:PIdent} for details) to determine confidence intervals for each parameter under 11 (blue dotted), 21 (red dashes), and 51 (green line) equispaced time points. 
$\bar\ell_p$ is $\chi^2$ distributed, thus values greater than $\bar\ell_p$ (displayed as a horizontal yellow line) all have a 95\% probability of being responsible for the synthetic dataset from which they are measured.
Thus, the points where the blue, red, and green curves intersect this line represent the 95\% confidence interval bounds of this estimation.
We normalise using the nominal values on the $x$-axis in each plot to allow comparison of the relative widths between parameters.
For all four parameters, the confidence intervals always contain the nominal value, and increasing the number of time points sampled consistently reduces the width of the confidence intervals, indicating improvement in measuring parameters.
For the green curve (51 data points), the peak value and confidence intervals give a very tight interval around the nominal value for $\chi_{PB}^{max}$, \Cref{fig:figure2}A. 
Parameters $a$ and $\chi_{BP}$, displayed in \Cref{fig:figure2}C and \Cref{fig:figure2}D, respectively, are relatively close ($\pm\sim10\%$) to the nominal value. 
The range of values on the x-axis is extended for $\chi_{PB}^{min}$ (\Cref{fig:figure2}B) to allow the confidence interval of the green curves to be measured (white and grey regions are given for visual reference to compare to the other plot's ranges).
The sparser datasets (11 and 21 measurements) in \Cref{fig:figure2}A and \Cref{fig:figure2}C produce reasonable estimated value measurements, with only \Cref{fig:figure2}A giving a measurable interval in the range examined in 11 measurement datasets. 
In \Cref{fig:figure2}B, the sparser datasets do not generate confidence intervals despite the larger domain, while the dense dataset can eventually give a value on a much larger range.

\subsection{Applying Genetic Algorithms to produce optimal sampling schedules} \label{res:genalgo}

\begin{figure}[ht]
\centering
    \includegraphics[width=\linewidth]{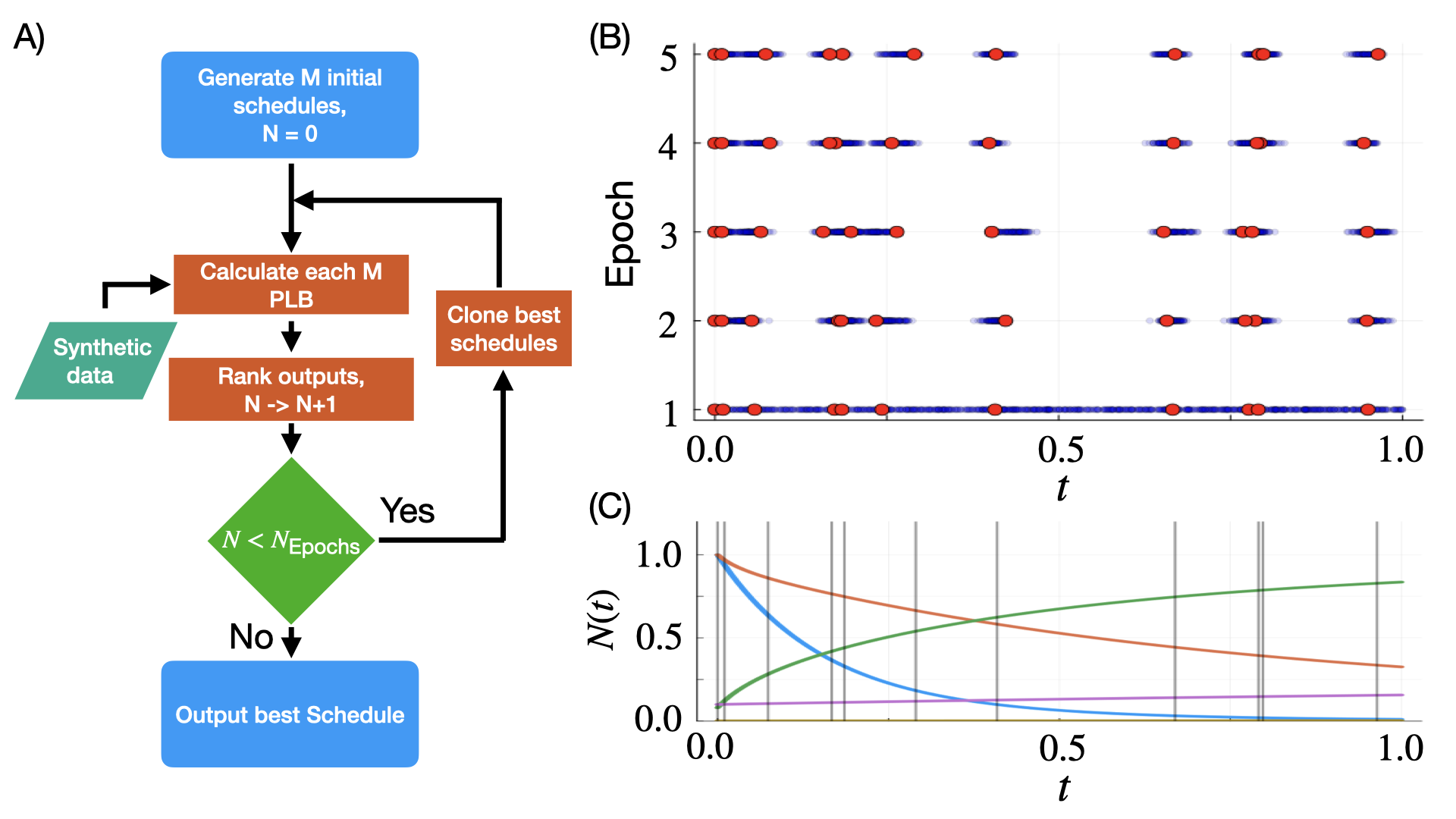}
    \caption{\textbf{Genetic algorithm schematic and examples.}
    (A) A flowchart highlighting the key steps of the genetic algorithm.
    Initially, 100 random schedules for the corresponding time points (11, 21, or 51) are fed in. 
    For each, a synthetic dataset with three biological replicates is generated according to the schedule, and the $95\%$ confidence interval width is found for the parameter of interest.
    The schedules are then ranked by width, with the smallest widths ranking more highly. 
    The process is repeated until the desired number of epochs has passed, with the best schedules being cloned between epochs (see section \Cref{SubSect:GenAlgo}).
    Finally, the best schedule after the desired number of epochs is output.
    (B) An example of sample schedule evolution through multiple epochs is shown. 
    The schedules are displayed as a low-opacity scatter plot for each epoch, with darker points indicating those that appear in multiple schedules. 
    The best schedule within the epoch is shown as a set of larger red points.
    As the epochs pass (i.e., moving up the lines on the plot), these points move diffusively as they are cloned.
    (C) A plot of the dynamics in the system is shown.
    Here, $N(t)$ is the population size of each compartment in the model at time $t$.
    For the best schedule (i.e., epoch 5 in (B)), the vertical lines shown represent the times at which sampling is optimal.
    }
    \label{fig:figure3}
\end{figure}

Our previous results indicate that dense sampling is needed to identify parameters accurately when using uniform sampling. 
We now investigate whether alternative sampling schedules can improve parameter identifiability. 
We use a genetic algorithm to systematically evaluate numerous schedules and determine their corresponding confidence interval widths (see \Cref{SubSect:GenAlgo}).
A flowchart of the algorithm, along with examples, is presented in \Cref{fig:figure3}. 
Keeping the same total number of data points as in \Cref{fig:figure2}, a corresponding optimal schedule was produced, using three replicates in each synthetic dataset, based on its ability to minimise the corresponding parameter's 95\% confidence interval. 

\begin{figure}[ht]
\centering
    \includegraphics[width=\linewidth]{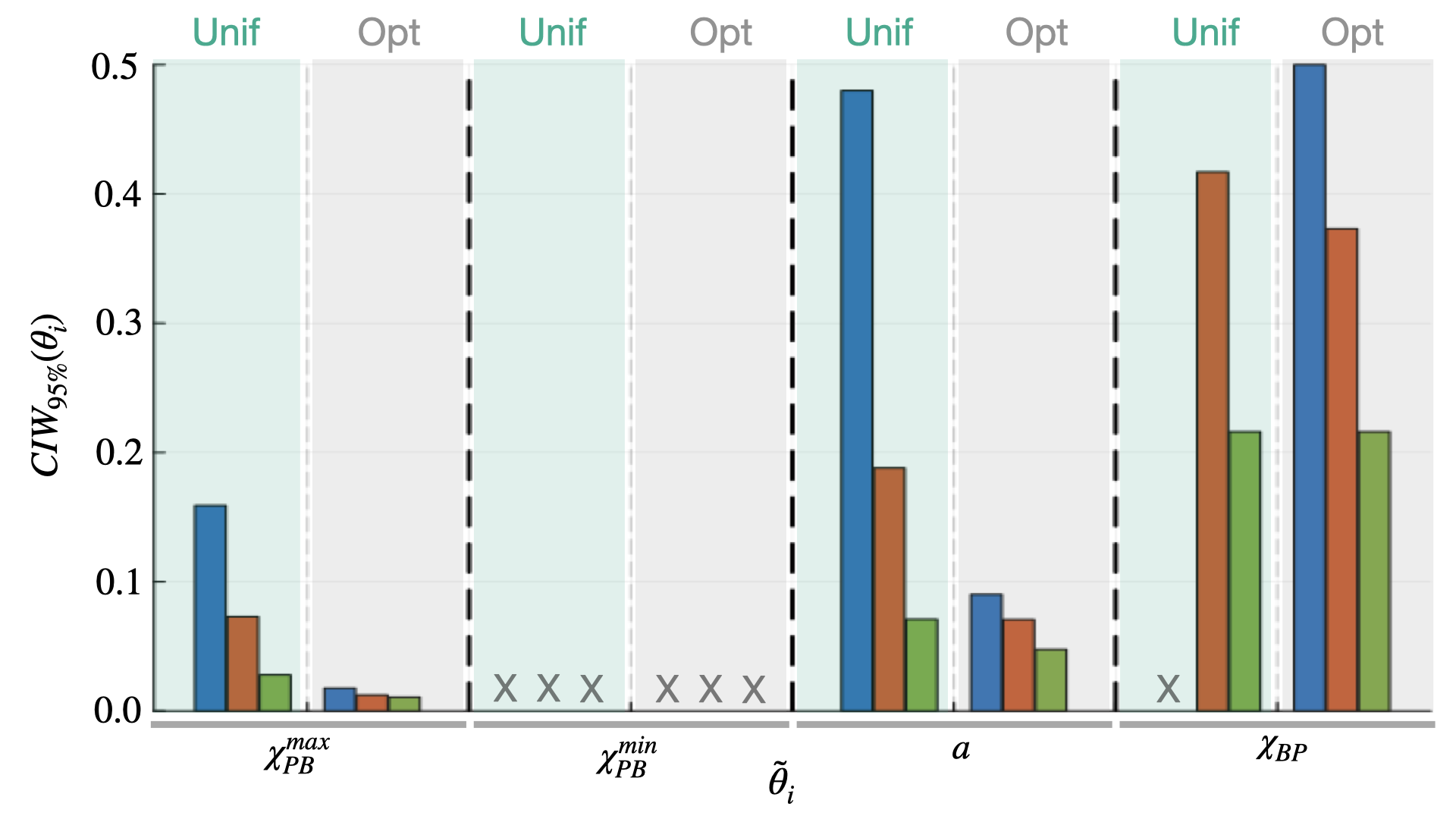}
    \caption{\textbf{ $95\%$ confidence interval width for different sampling schedules and number of time-points.}
    Each parameter of interest has two sets of values, separated by light grey dashed lines, with parameters separated by dark black dashed vertical lines.
    For a given parameter, uniform time-point sampling values are given (light green columns), and optimal sampling from the genetic algorithm (light grey columns).
    The level of data with sampling types is 11, 21, and 51 time points, with blue, red, and green bars, respectively. 
    Where no value could be found (i.e., $CIW_{95\%}>0.7$), a grey X is given instead of a bar.
    }
    \label{fig:figure4}
\end{figure}

\begin{table}[!ht]
\centering
\begin{tabular}{|c||c|c|c||c|c|c|}
\hline
Parameter         & U-11 & U-21 & U-51 & Op-11 & Op-21 & Op-51  \\
\hline \hline
$\chi_{PB}^{\max}$ & 0.159 & 0.073 & 0.028 & 0.0177 & 0.0123 & 0.0107 \\
\hline
$\chi_{PB}^{\min}$ & NaN	& NaN & NaN & NaN & NaN & NaN \\
\hline
$a$               & 0.480 & 0.188 & 0.071 & 0.0901 & 0.0708 & 0.0473 \\
\hline
$\chi_{BP}$       & NaN	& 0.417 & 0.216 & 0.500 & 0.373 & 0.216 \\
\hline
\end{tabular}
\caption{
\textbf{Parameter confidence intervals for uniform and optimal sampling strategies and levels of data.} 
The 95\% confidence interval widths are given for each parameter estimation.
The columns titled U-N apply a uniform sampling schedule consisting of $N$ sample points, including the initial condition.
The Op-N columns have the genetic algorithm-optimised sampling schedules for $N$ sample points, including the initial condition.
Where no value could be estimated on the interval, the value NaN is given.
The search interval was extended (see \Cref{fig:figure2}B, where the grey region indicates the extended range) to quantify a value in the following best-case scenario where possible (e.g., 51 sampling times for $\chi_{PB}^{\min}$). 
The values are also visualised in \Cref{fig:figure4}.
}
\label{tab:ConfIntervalWidths}
\end{table}

We compare the relative confidence interval widths between uniform and optimal sampling in \Cref{fig:figure4} and \Cref{tab:ConfIntervalWidths}. 
For each parameter (separated by black dashed lines), there is a section representing uniform sampling (green background) and optimal sampling (grey background). 
Within each section, blue bars represent 11 time points, orange bars represent 21 time points, and green bars represent 51 time points. 
X's represent a confidence interval larger than 0.7.
As with uniform schedules, an increase in the available data (e.g., the number of time points) reduces the confidence interval, resulting in a more accurate parameter estimate.
The trend in identifiability of the parameters is similar to uniform sampling, with $\chi_{PB}^{\max}$ and $a$ showing the smallest interval widths, followed by $\chi_{BP}$, and then $\chi_{PB}^{\min}$, which was not identifiable under any circumstances.
By optimally choosing our sample schedules, we achieve a confidence interval width for 11 optimally spaced measurements that is comparable to, and in some cases narrower than, that of 51 uniformly spaced measurements.
However, there are diminishing returns for optimal sampling compared to the improvement seen in the uniform sampling case.
For example, when comparing $\chi_{PB}^{\max}$ measured with 11 vs. 51 sampling points, the confidence interval for uniform and optimal sampling was roughly one-sixth and one-half as wide, respectively. 
Even with optimal sampling, $\chi_{PB}^{\min}$ was highly non-identifiable, even in the highest data availability scenarios. 

We observe stark differences in the identifiability of the parameters given these data scenarios.
Overall, the maximal speed of binding $\chi_{PB}^{\max}$ is most readily measured and has the lowest associated uncertainty in all scenarios, suggesting that its value is essential in the overall dynamics of the system. 
The transition between fast $\chi_{PB}^{\max}$ and slow $\chi_{PB}^{\min}$ binding, governed by the value of $a$, as well as the binding off rate $\chi_{BP}$, were the following most readily measured parameters.
This suggests they have some impact on the overall dynamics, but a lesser one than the maximal binding rate.
Finally, we saw that the slow binding rate $\chi_{PB}^{\min}$ was not practically identifiable in all of the data scenarios.
In contrast to $\chi_{PB}^{\max}$, the specific value didn't readily impact the overall likelihoods, likely because the values were all small.
As such, very few bacteria bind at this low rate, and so moderate changes to this tiny population are not easy to distinguish in light of the faster modifications dictated by the other parameters.

\subsection{Characterising biofilm size sensitivity to formation parameters} \label{res:sena}

\begin{figure}[ht]
\centering
    \includegraphics[width=\linewidth]{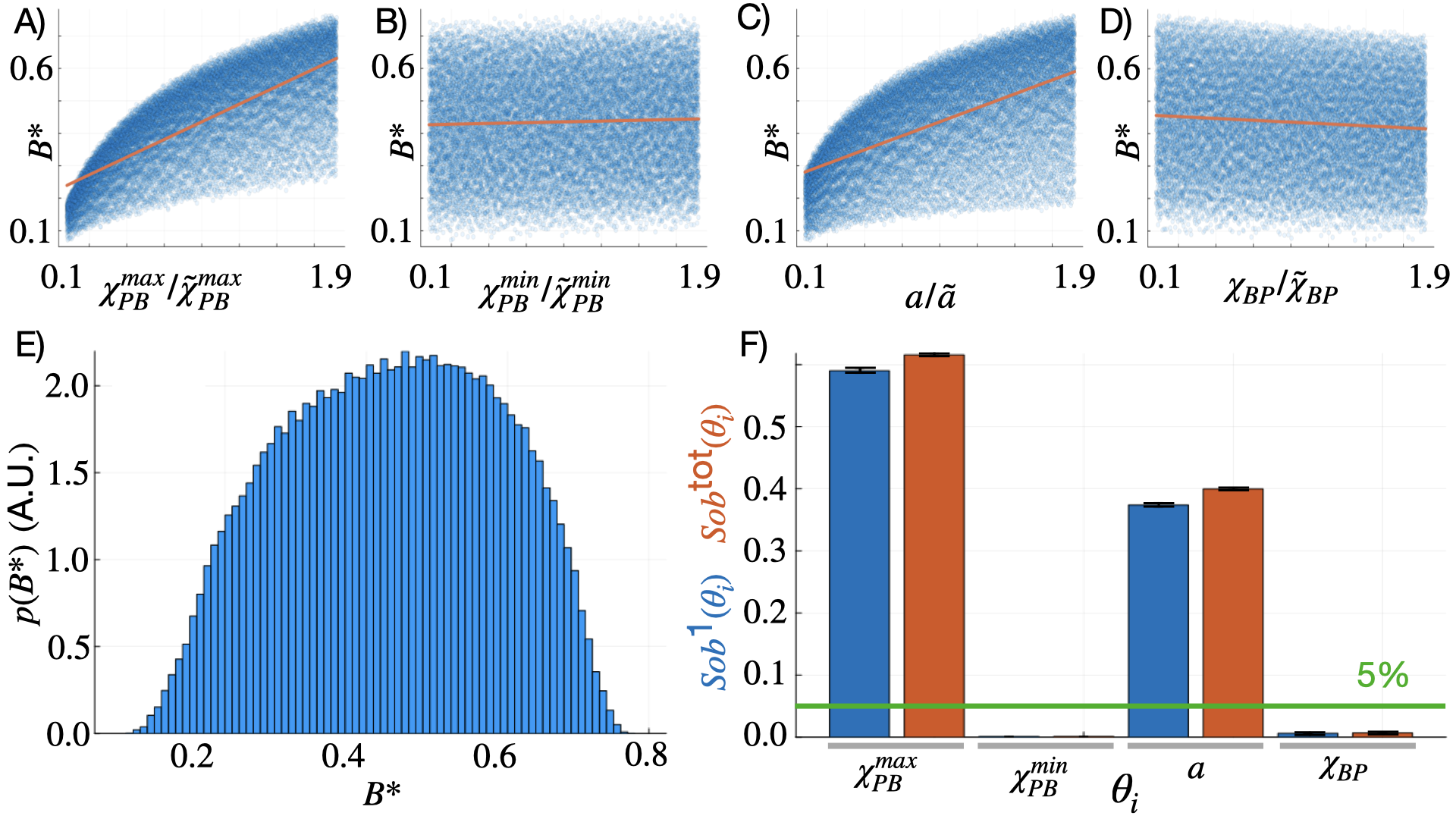}
    \caption{\textbf{Sensitivity analysis of the end-point biofilm size given variations in binding dynamics.}
    (A-D) Scatter plots of final biofilm size for parameters $\chi_{PB}^{max}$ (A), $\chi_{PB}^{min}$, (B), $a$ (C), and $\chi_{BP}$ (D) over 40,000 Saltelli sampling parameter sets.
    For each scatter plot, a linear fit (red solid line) illustrates general trends in the data, and corresponding fit and correlation values are provided in \Cref{tab:senIndex}.
    (E) The distribution of the end-point biofilm sizes across the parameter space, sampled from (A-D), is given.
    (F) The variance presented in (E) is decomposed to give the associated first-order (blue) and total-order (orange) Sobol sensitivity indices.
    Bars above the green horizontal line, representing the 5\% threshold, indicate parameters that impact overall biofilm size.
    }
    \label{fig:figure5}
\end{figure}

Having examined the identifiability of the different parameters, we next turn towards sensitivity analysis to determine the impact of varying these parameters on the final size of the biofilm, denoted $B^*$ (see details in \Cref{SubSect:SenA_1}).
\Cref{fig:figure5}A-D displays the univariate dependence of $B^*$ on the normalised underlying parameter.
Parameters are normalised to allow for a comparison of how proportional changes affect the output.

\begin{table}[!ht]
\centering
\begin{tabular}{|l||c|c|c|c|}
\hline
Index type & $\chi_{PB}^{\max}$ & $\chi_{PB}^{\min}$ & $a$ & $\chi_{BP}$ \\
\hline\hline
Intercept    & 0.218 &  0.426  & 0.263   &  0.460  \\
Gradient     & 0.218 &  0.010  & 0.172   & -0.023  \\
\hline
Pearson      & 0.749 &  0.034  & 0.592   & -0.078  \\
Partial      & 0.935 &  0.118  & 0.901   & -0.266  \\
\hline
Spearman CC  & 0.014 & 0.005  & -0.002  & -0.001  \\
PRCC         & 0.014 & 0.010  &  0.001  & 0.000  \\
\hline
Sobol tot    & 0.616 &  0.001  &  0.400  &  0.007  \\
Sobol 1st    & 0.590 &  0.001  &  0.374  &  0.006  \\
\hline
\end{tabular}
\caption{
\textbf{Various sensitivity indices quantifying impacts on $B^*$ for select phenotype transition parameters.} 
The table shows several metrics to quantify the dependence of the final biofilm size, $B^*$, on the model's parameters.
}
\label{tab:senIndex}
\end{table}

\Cref{tab:senIndex} depicts several sensitivity indices of the univariate relationships captured in \Cref{fig:figure5}A-D.
The strongest linear relationships, measured by either the gradient of the linear fit or the Pearson correlation coefficient, are present in $\chi_{PB}^{\max}$ and $a$.
The remaining parameters only have weak linear dependence. 
The Partial Correlation for $\chi_{PB}^{\max}$ and $a$ reinforces our observation, providing a clearer relationship when controlling for the other parameters.
Interestingly, the magnitude of the Partial Correlation for $\chi_{PB}^{\min}$ and $\chi_{BP}$ increases in both cases, suggesting the influence of these parameters is being suppressed by variation in other parameters.
In all cases, the Spearman and Partial Rank correlations (PRCC) are small, indicating that outliers primarily drive the linear relationships suggested by our previous metrics.
In \Cref{fig:figure5}E, we can see this distribution of values, with the bulk of the values spread across $B^*=(0.2,0.8)$.
The mean across the whole sample is $E(B^*)=0.435$, which is not far from the nominal value $ B^*_0 = 0.484$.
Finally, to decompose the relative impact of the parameters on $B^*$ regardless of their underlying relationship, we calculated the Sobol indices via a variance-based decomposition (\Cref{fig:figure5}F).
Supporting what has been seen up to this point, $\chi_{PB}^{\max}$ and $a$ are dominant, accounting for 0.95 of the total variance.
Additionally, we can see that their total-order Sobol index is very close to their first-order Sobol index.
Thus, the influence of these two parameters has an independent impact on the quantity of interest.

Following our concluding remarks from \Cref{res:genalgo}, the results seen here further suggest that the overall dynamics of the bacteria are strongly impacted by the value of $\chi_{PB}^{\max}$.
That is, with planktonic bacteria quickly able to bind to biofilms, the result is a significantly larger difference in final biofilm size $B^*$.
The value of $a$ also has a relatively strong impact on this final biofilm size, likely since this value sets the duration for which the $\chi_{PB}^{\max}$ rate dictates the planktonic bacteria binding.
As such, we see that bacteria that can maintain their fast-binding dynamics result in far larger final sizes $B^*$.
The values of $\chi_{PB}^{\min}$ and $\chi_{BP}$ have relatively little impact on the final biofilm size, likely because their values are small. Therefore, bacteria that detach more readily or attach to large biofilms (i.e., $B >> a$) have relatively little impact on the overall final size of the biofilm.

\subsection{Capturing bacterial evolution by selection with structured models} \label{res:struct}

\begin{figure}[htb!]
\centering
    \includegraphics[width=0.95\linewidth]{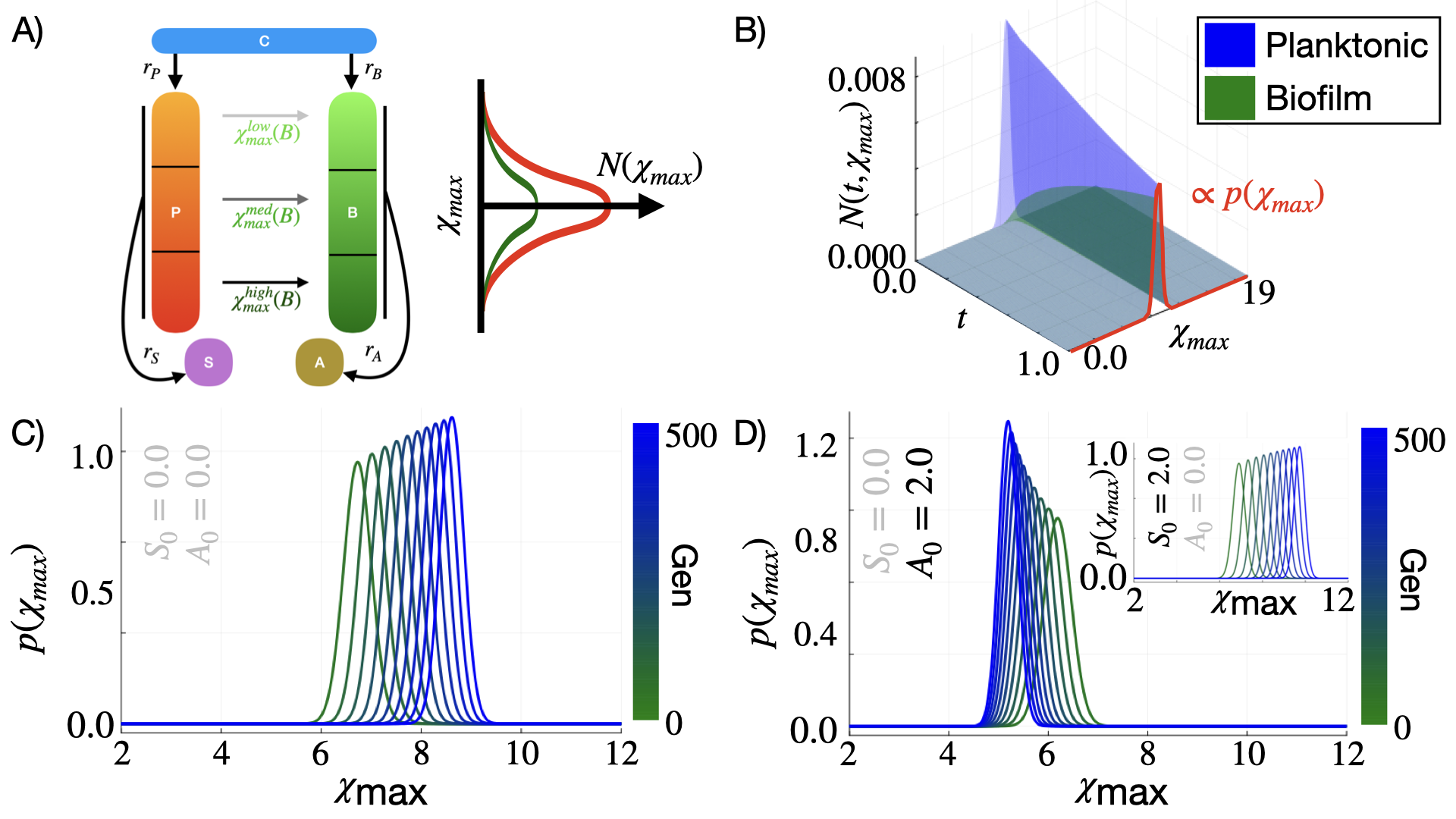}
    \caption{\textbf{Schematics, examples, and long-time dynamics of the phenotype parameter-structured model.}
    (A) A schematic of the parameter-structured population model (see \Cref{Sect:Model2}).
    The key compartments are as in the short-term model: a media carbon source $C$, planktonic bacteria $P$, biofilm bacteria $B$, planktonic predator $S$, and biofilm predator $A$.
    Within the phenotype compartments, sub-communities are separated based on their ability to attach to the biofilm when in the planktonic phenotype.
    Within the populations $P$ and $B$, the distribution of organisms with associated $\chi_{max}$ is normally distributed around some value $\bar\chi_{max}$.
    (B) An example evolution showing all of the compartments (here, characterised by differences in their $\chi_{max}$ value).
    The blue surface shows the evolution of the planktonic compartments, while the green surface shows the evolution of the biofilm compartments.
    The height $N(t,\chi_{max})$ gives the population at time $t$ within the population compartment $\chi_{max}$.
    The final time-point distribution of the biofilm's $\chi_{max}$ values, $p(\chi_{max})$, is highlighted in red.
    (C) The final time-point distribution of $\chi_{max}$ within the biofilm is presented for 500 model generations.
    In this example, the model has no predation (i.e., $S_0$ and $A_0$ are zero throughout).
    The early curves are shown in green, while the later ones are given in blue. 
    The curves can be seen moving slowly to the right as the generations increase.
    (D) The final time-point distribution of $\chi_{max}$ within the biofilm is presented for 500 model generations.
    The main figure models a population subject to a periodically reset population of biofilm predators.
    Here, the $\chi_{max}$ distribution drifts to the left.
    The inset models a population with a periodically reset planktonic predator.
    Here, the $\chi_{max}$ distribution drifts to the right.
    }
    \label{fig:figure6}
\end{figure}

After evaluating the parameters that determine biofilm formation in the short term, we propose a structured model that includes a mechanism for long-term population changes through evolution (see \Cref{Sect:Model2}). 
The key difference in our new model is that our system has multiple bacterial populations, each with distinct binding dynamics. 
An example schematic is presented in \Cref{fig:figure6}A. 
This example features three compartments with different maximal binding rates, the values of which in the previous model were denoted $\chi_{PB}^{\max}$.
Based on the results of the identifiability and sensitivity studies of the previous model, we removed the parameters $\chi_{BP}$ and $\chi_{PB}^{\min}$, since their impact on the system was minimal. 
Additionally, instead of a scalar initial condition for $P$ and $B$, the new populations $P_i$ and $B_i$ are given some initial distribution, namely normally distributed across some range of $\chi_{PB}^{\max}$ values, with the $i$th compartment having a value $\chi_i$. 

As detailed in \Cref{Sect:Model2}, we no longer focus only on simulations of single time intervals from $t=0$ to $t=1$, but instead take the state of the system at $t=1$ to reset the system periodically.
In this way, it simulates multiple generations of the population subjected to a consistent and resetting environment. 
\Cref{fig:figure6}B presents an example of the time interval $t=0$ to $t=1$ for these compartment populations.
At $t=1$, our biofilm population is highlighted by a red curve; this population is used to reseed the planktonic population in the next generation.

We demonstrate the presence of selection by measuring the average binding rate $\bar\chi_{\max}$ present in our model both under no predation (\Cref{fig:figure6}C) and with predation (\Cref{fig:figure6}D). 
These figures display the probability distribution of the compartments $\chi_{\max}$ values. 
Without predators, the distribution shifts to the right as the system undergoes many generations and narrows.
This right shift indicates that bacteria with a higher $\chi_{\max}$ value (i.e., those that more readily attach to the biofilm) are more represented at the end of each generation and are therefore being reseeded into the next generation.
The increased peak height suggests that the distribution is narrowing over time, slowing the rate at which the curve moves to the right. 
Under no predation, we are naturally selecting for bacteria that can quickly attach to the biofilm, as these are the ones that make it to the next generation.
The evolutionary effects of predator stress are displayed in \Cref{fig:figure6}D.
The main figure features biofilm predators, and the inset figure has planktonic predators, both of which are periodically refreshed.
The biofilm predator drives opposite selection from the no-predator case, with the curve shifting to the left and favouring smaller values of $\chi_{\max}$ where bacteria transition to the biofilm more slowly.
As before, this curve is narrowing, indicating that the process slows down as the bacteria undergo more generations.
The planktonic predator dynamics look similar to the no-predator case, with bacteria with a higher $\chi_{\max}$ being favoured; however, it shifts right at a faster rate.
The presence of predators results in very different types of selection within this model, which drives different outcomes in the bacterial population.

\subsection{Characterising evolutionary dynamics' sensitivity} \label{res:sena2}

\begin{figure}[ht]
\centering
    \includegraphics[width=\linewidth]{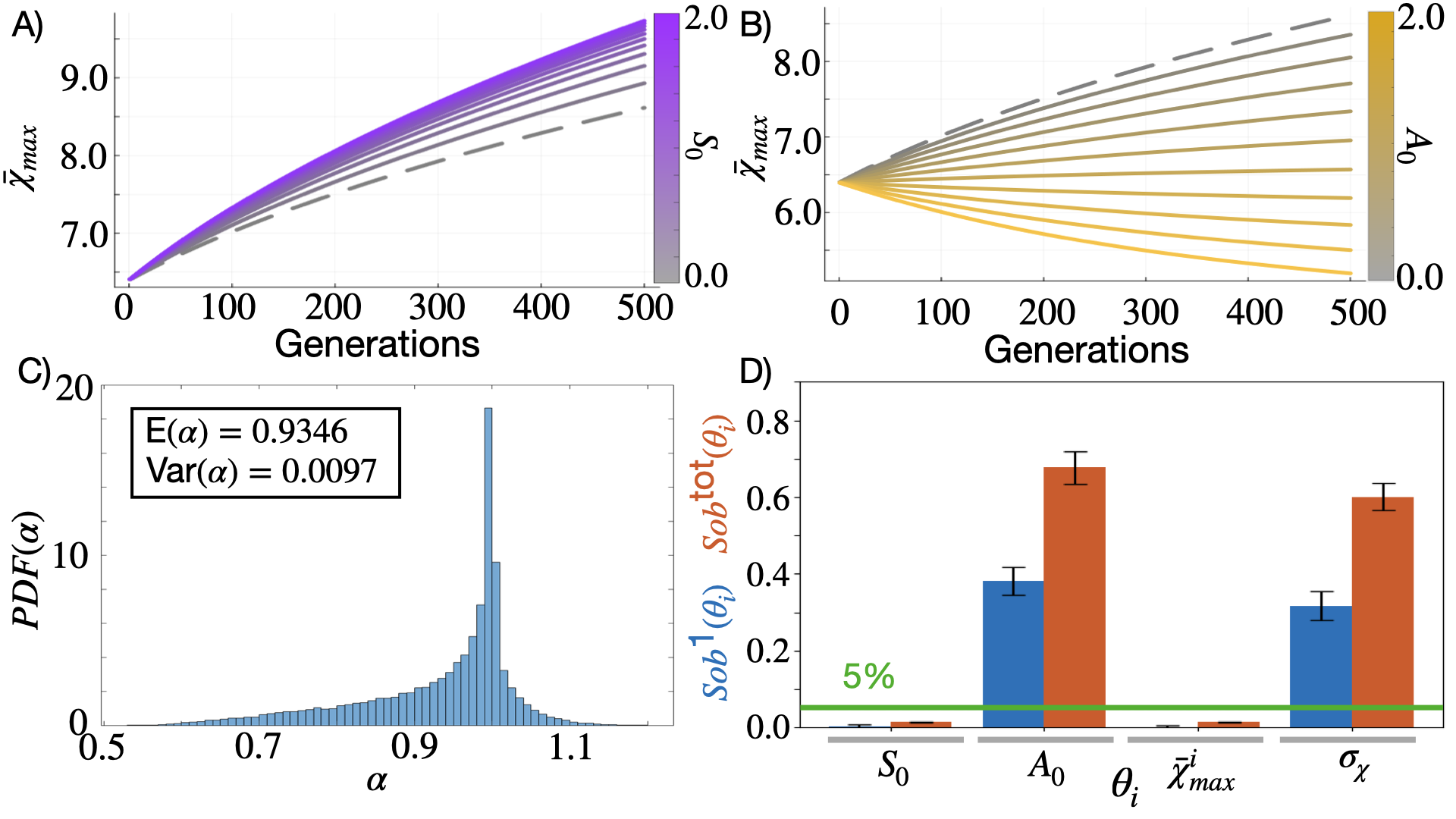}
    \caption{\textbf{Characterising trends within the $\chi_{max }$ distributions across generations and measuring the sensitivity of these trends to conditions.}
    The biofilm population average, $\bar\chi_{max}$, for 10 varying amounts of resetting predator populations for planktonic (A) and biofilm (B) predators over 500 generations. 
    The colour bar provides the size of the resetting predator population for each model.
    (C) The distribution of the quantity of interest   $\alpha=\left(\frac{\bar\chi_{max}^f}{\bar\chi_{max}^i}\right)$ at 50 generations.
    The inset gives the mean and the variance of the Saltelli samples.
    (D) The variance presented in (C) is decomposed to give the associated first-order (blue) and total-order (orange) Sobol sensitivity indices.
    }
    \label{fig:figure7}
\end{figure}

To further understand the relationship between predation and bacterial populations, we can examine how the number of predators within the system affects their evolution.
In \Cref{fig:figure7}A and \Cref{fig:figure7}B, we measure the average binding rate in the population, $\bar\chi_{\max}$, at the end of each generation. 
\Cref{fig:figure7}A shows that over 500 generations, the population subjected to planktonic predators tends toward higher values of $\bar\chi_{\max}$, indicating that biofilms are forming quicker.
These predators compound the selection effects in the no-predator system, further increasing the value of $\bar\chi_{\max}$ in the population.
In contrast, \Cref{fig:figure7}B shows a stronger relationship between the predator presence and the binding rate.
Interestingly, \Cref{fig:figure7}B transitions from an increasing $\bar\chi_{\max}$ value to a decreasing one as the predator quantity increases, as the biofilm predators overcome the selection from the experimental design.
All the curves in \Cref{fig:figure7}A and \Cref{fig:figure7}B show varying slowing levels, with the gradient over time decreasing, suggesting each predator level may drive some asymptotic binding value on very-long timescales. 

To understand the most significant drivers in selection for biofilm formation, we undertook a sensitivity study on the structured model (described in \Cref{SubSect:SenA_2}).
We examined the effects of (1) the reset population of the planktonic predator $S_0$, (2) the biofilm predator $A_0$, (3) the initial population distribution mean $\bar\chi_{\max}$, and (4) the population initial standard deviation $\sigma_\chi$.
The quantity of interest measured for our sampling was the ratio of the final population average binding rate to the initial.
We used measurements at 50 generations based on the results presented in \Cref{fig:figure7}A and B.
These figures clearly distinguish their population's average binding rates by 50 generations.
Since these simulations are relatively computationally expensive, this choice balances execution times with simulations that are long enough to allow a clear distinction to result from the selection impacts.
We examined a sample across the parameter space using the Sobol sampling method.
\Cref{fig:figure7}C displays the observed values distribution.
We can see that the distribution is skewed to the left, with most of the samples reducing the final average binding rate compared to the initial. 
After measuring the overall variance in this distribution, we calculated the Sobol sensitivity indices to determine which parameters drive these variances (\Cref{fig:figure7}D).
Most of the variation is attributed to the variance of the parameters $A_0$ and $\sigma_\chi$, the resetting biofilm predator population size, and the initial bacterial binding variance, respectively.
Unlike our previous sensitivity analysis, the first-order and total Sobol indices were unequal, indicating that the parameters are no longer independent.
Instead, higher-order effects are also at play.
The second-order indices have no significant effect, except the index between $A_0$ and $\sigma_\chi$, which had a value $0.274 \pm 0.055$. 
Given first-order index values of $A_0$ and $\sigma_\chi$ of 0.381 and 0.317, respectively, the combination of these first-order values and the measured second-order between them explains $97.2\%$ of the overall variance in the data.
This suggests that populations with solely high parameter variance, or solely under high levels of predation, are insufficient to drive the most significant levels of change.
Instead, these two factors cooperate, driving the most considerable deviations from the nominal.

These results suggest that bacterial populations subject to such biofilm predators should be expected to display signs of selection in their biofilm formation dynamics most rapidly.
While planktonic predation will drive some changes in biofilm formation dynamics, these changes occur more slowly.
For less diverse populations of bacteria, as is common in lab strains, this lack of diversity likely limits the rates of selection since there is limited capacity of sub-populations to outperform others.

\section{Conclusions} \label{sec:conc}

This study investigated models of bacteria switching between planktonic and biofilm-forming phenotypes, with results presented in two main areas. 
First, we explored the short-term population dynamics of bacteria under predation stress, assessing the real-world applicability of a mathematical model through structural and practical identifiability, optimal design, and sensitivity analyses. 
Recognising the short-term model's limitation in reflecting long-term population changes, we then proposed a new structured long-term model. 
Our minimal approach divided phenotypic populations into compartments with varying behaviours, allowing some to outperform others. 
By periodically resetting our system and preserving the population distribution of this transition behaviour, mirroring the Long-Term Evolution Experiment (LTEE) design often conducted in laboratory settings, we enabled selection to occur within our model.

In the first part of our study, we analysed an existing stoichiometric model that describes bacteria transitioning between planktonic and biofilm phenotypes and their associated predators.
We first confirmed the system's global structural identifiability when information from all state variables was available. 
Furthermore, we demonstrated that restricting the parameters to those involved in biofilm dynamics ($\chi_{PB}^{\max}$, a, $\chi_{PB}^{\min}$, and $\chi_{BP}$), only information about the biofilm state variable ($B$), is required for global structural identifiability.

We then generated synthetic datasets from numerical solutions to estimate these biofilm dynamics parameters.
We used Maximum Likelihood Estimation to estimate parameters from synthetic data. 
We found that, while the fitted solutions accurately reproduced the overall trends in the data, our practical identifiability study revealed that the accuracy of parameter reconstruction varied with sampling frequency and scheduling. 
In a best-case scenario, across a low-noise scenario with 11 to 51 measurements, the maximal binding rate of planktonic cells ($\chi_{PB}^{\max}$) and the biofilm size at which the binding rate decreased ($a$) exhibited the narrowest confidence intervals. 
This was likely due to initial conditions featuring a large planktonic population and a small biofilm, as is common in experimental setups. 
Conversely, $\chi_{PB}^{\min}$ and $\chi_{BP}$ were more challenging to measure accurately, likely because of their limited activity periods and smaller magnitudes.
We propose that modifying experimental protocols, such as pre-growing biofilms, could enhance the measurement of these less identifiable parameters by shifting initial conditions to emphasise their roles in the dynamics. 
However, this approach might compromise the accuracy of other parameter measurements. 
Finally, our application of a genetic algorithm revealed that optimised sampling schedules could achieve data quality comparable to, or even exceeding, that of significantly higher uniform sampling rates, suggesting a limited inherent informativeness of state variables regarding underlying parameters.

Also, we conducted a sensitivity analysis on the initial model, which clearly demonstrated that $\chi_{PB}^{\max}$ and $a$ dominated the system's overall variance. 
These results, combined with our parameter identifiability findings, led us to simplify the model for initially small biofilms. 
Thus, in our long-term structured model (Section \eqref{Sect:Model2}), we omitted the biofilm-to-planktonic transition rate ($\chi_{BP}$) and the large-biofilm binding rate ($\chi_{PB}^{\min}$).
Our long-term model organises planktonic and biofilm populations into compartments, each defined by a unique maximum transition rate ($\chi_{PB}^{\max}$) from planktonic to biofilm. 
Initially, $\chi_{PB}^{\max}$ values are normally distributed within the planktonic bacteria. 
While the short-term evolution of this system qualitatively resembles the initial model, periodic reseeding using the biofilm's $\chi_{PB}^{\max}$ distribution generates emergent differences over many generations. 
Selection was demonstrated to occur even in the absence of predators; this is likely driven by the reseeding mechanism, which acts as a selective pressure.

We observed that the presence of predators significantly alters the dynamics of selection. 
The biofilm-specific predator ($A$) had a significantly greater impact, resulting in faster adaptation. 
In contrast, a planktonic predator ($S$) had a much smaller effect compared to the inherent selection in the no-predator scenario. 
Interestingly, while the biofilm predator strongly influenced adaptation, the rate of adaptation slowed down more rapidly than in the other cases. 
This suggests a trade-off: organisms must transition quickly enough to form biofilm by the end of the generation, but not so quickly that they risk prolonged exposure and consumption within the biofilm. 
Consequently, the distribution of $\chi_{\max}$ shifts towards an optimal value that balances this trade-off.

We performed sensitivity analysis to understand how the ratio between final and initial mean binding rates is impacted by key model inputs: initial planktonic predator population size ($S_0$), initial biofilm predator population size ($A_0$), initial bacterial mean transition rate ($\bar\chi_{\max}$), and initial distribution mean ($\sigma_\chi$).
This analysis revealed that variance is dominated by biofilm predator ($A_0$) and initial system variance ($\sigma_\chi$). 
The interaction between $A_0$ and $\sigma_\chi$ accounted for most of the unexplained variance. 
This is because $A_0$ directly impacts the biofilm population, exerting a strong and clear selective pressure. 
Similarly, a larger $\sigma_\chi$ enhances the system's capacity to respond rapidly to selection, as it provides a greater initial diversity of competitively advantageous populations.

Our current models offer several avenues for future exploration. 
Although a comprehensive investigation of genetic algorithms was beyond the scope of this study, we found that optimal sample points often clustered. 
Future research could address this by either removing these clustered points or replacing them with new, randomly drawn points that maintain a minimum pairwise distance. 
Additionally, minimising the sum of all confidence intervals, as suggested by \parencite{lam_evolving_2024}, could yield more robust sampling schedules. 
Finally, data sets incorporating multiple initial conditions could be employed to yield more accurate parameter set measurements, especially in models with regime-dependent identifiability.

Our structured model effectively captures selection dynamics; however, there is a need for further refinement to reflect real-world complexities, particularly the inherent limitations in bacterial cellular processes. 
Currently, our model simplifies these by characterising compartment differences solely based on $\chi_{\max}$ values, which could represent variations in attachment ability. 
However, selection likely also involves other parameters, such as bacterial reproduction rates ($r_P$ and $r_B$). 
Furthermore, real biological systems may impose discrete or bounded constraints on $\chi_{\max}$ values, and multiple underlying parameters often vary simultaneously under stress, potentially incurring energetic costs for traits like increased $\chi_{\max}$. 
Our model also assumed that all possible $\chi_{\max}$ values were initially present. 
To create more realistic models of systems, future studies should explore constrained and simultaneous parameter changes, incorporate mutation dynamics between compartments, and crucially, align these model additions with experimental observations to avoid unjustified theoretical assumptions.

\section*{Statements and Declarations}

\textbf{Conflict of Interest:} The authors declare that they have no known competing financial interests or personal relationships that could have appeared to influence the work reported in this paper.\\
\textbf{Data Availability:} All data generated and codes used in analysis during this study are included in the published article's associated GitHub at https://github.com/StephWilleniams/BiofilmModelAnalysis

\section*{Author Contributions} 

All authors contributed to the study conception and design.
Stephen Williams performed numerics, data collection, and analysis. 
All authors read and approved the final manuscript.

\section*{Acknowledgements} This material is based upon work supported by the National Science Foundation under Grant
No DBI-2214038.

\printbibliography

\section{Supplementary Materials}

\subsection{Nondimensionalisation of the Model} \label{SubSect:NDprocess}

In the original model, as presented in \parencite{seiler_grazing_2017}, the carbon, planktonic, and planktonic predators have units of carbon per volume.
The biofilm and its associated predator have units of carbon per area.
These units are generally convenient, giving the closest description to what would be measured experimentally. 
However, the mixed-compartment units in the model can provide a somewhat obscured comparison of the relative sizes of the populations, making them challenging to compare.
As such, the equations describing the populations were multiplied by their associated dimensions to avoid inconsistent comparisons, either a volume of $3\mu$l or an area of $8$cm$^2$.
With this change, the populations represent the total quantity of carbon stored in each state variable.
For consistency, several parameters must also be handled this way, for example, $H_C$.
Upon this dimensionalisation, the area/volume value in the original model equations was absorbed into other parameter values, where necessary.
As such, the values of $\chi_{PB}^{\max}$ and $\chi_{PB}^{\min}$ have been scaled by the original system's area/volume ratio, and $a$ by the system area.
In addition, the parameters and timescales were also normalised so that 1 time unit corresponds to 1 day, since we entirely focus on simulated runs of this length.
The efficiency calculation is performed to enforce model structure consistency, and the values of $r_p$ and $ r_b$ are scaled by $1/e_b$. 
The efficiency value $e_b$ is applied in Equations \eqref{eqn:ODE_P} and \eqref{eqn:ODE_B}, to be consistent with how it is used for Equations \eqref{eqn:ODE_S} and \eqref{eqn:ODE_A}.

We choose our initial conditions for the system to be $C=1$ and $P=1$, corresponding to $0.3\dot{3}\mu$gC/ml in the dimensionalised units; the remaining populations are then chosen to be some orders of magnitude smaller.
This choice provides a convenient way to consider how these populations evolve relative to their initial condition, which becomes an essential consideration in \Cref{Sect:Model2} to facilitate a simple reseeding process.

A limitation of this nondimensionalization approach is that the system's parameter values are then tailored to the specific geometry, since the underlying parameters in \parencite{seiler_grazing_2017} were scaled to fit the given experimental container. 
Another potential approach was scaling the state variables and parameters to the initial carbon content sum across compartments. 
This normalisation would allow a convenient way to view how the carbon is distributed proportionally.
Additionally, that would clarify what carbon is lost through efficiency ($e_b$, $e_S$, $e_A$).
However, the initial conditions favouring scaling were implemented, since we are primarily interested in between-parameter comparisons.

\begin{table}[t!]
    \centering
    \begin{tabular}{|c|c|c|c|c|}
        \hline
        Variable & Name & ND Value & Value & Units \\
        \hline\hline
        & Fixed parameters & & & \\
        \hline
        $e_b$ & Bacterial growth efficiency & 0.2 &  & \\
        \hline
        $r_P$ & Planktonic growth rate & 25.2 & 0.21 & h$^{-1}$  \\
        \hline
        $H_P$ & Planktonic Half-Saturation & 3.0 & 1 & $\mu\text{g ml}^{-1}$ \\
        \hline
        $r_B$ & Biofilm growth rate & 0.84 & 0.007 & h$^{-1}$  \\
        \hline
        $H_B$ & Biofilm Half-Saturation & 8.0 & 1.0 & $\mu\text{g ml}^{-1}$ \\
        \hline
        $g_S$ & Planktonic predator growth rate & 5.76 & 0.12 & $\mu\text{g ml}^{-1}$ \\
        \hline
        $e_S$ & Planktonic predator growth efficiency & 0.5 & &  \\
        \hline
        $H_S$ & Planktonic predator half-saturation & 3.0 & 1.0 & $\mu\text{g ml}^{-1}$ \\
        \hline
        $g_A$ & Biofilm predator growth rate & 4.32 & 0.09 & h$^{-1}$ \\
        \hline
        $e_A$ & Biofilm predator growth efficiency & 0.33 & &  \\
        \hline
        $H_A$ & Biofilm predator half-saturation & 0.8 & 0.1 & $\mu\text{g cm}^{-2}$ \\
        \hline
        \hline
        & Re-estimated parameters & & & \\
        \hline
        a & Attachment strength parameter & 0.08 & 0.01 & $\mu\text{g cm}^{-2}$\\
        \hline
        $\chi_{PB}^{\max}$ & Biofilm attachment rate maximum & 3.2 & 0.05 & cm h$^{-1}$ \\
        \hline
        $\chi_{PB}^{\min}$ & Biofilm attachment rate minimum & 0.032 & 0.0005 & cm h$^{-1}$ \\
        \hline
        $\chi_{BP}$ & Biofilm detachment rate & 0.12 & 0.005 & h$^{-1}$\\
        \hline\hline
        & Initial conditions & & & \\
        \hline
        $C(0)$ & Initial media concentration & 1.0 & 0.33 & $\mu$gC ml$^{-1}$\\
        \hline
        $P(0)$ & Initial plankton concentration & 1.0 & 0.33 & $\mu$gC ml$^{-1}$ \\
        \hline
        $B(0)$ & Initial biofilm concentration & 0.01 & 0.00125 & $\mu$gC cm$^{-2}$\\
        \hline
        $S(0)$ & Initial ciliate concentration & 0.03 & 0.01 & $\mu$gC ml$^{-1}$ \\
        \hline
        $A(0)$ & Initial amoeba concentration & 0.008 & 0.001 & $\mu$gC cm$^{-2}$ \\
        \hline
    \end{tabular}
    \caption{\textbf{Population model parameters comparsion to original values} Table of each of the parameters present in Equation \eqref{eqn:ODE_C}-\eqref{eqn:ODE_A}, and  \Cref{fig:figure1}A. 
    The values are a nondimensionalization of those presented in \parencite{seiler_grazing_2017}; see \Cref{SubSect:NDprocess} for details on this process.
    A brief description of each nominal non-dimensional value is given. 
    The parameter values, or our dimensionalised value equivalents, from \parencite{seiler_grazing_2017}, and their units are given in the final two columns.
    The first set of parameters is fixed and well-known within our model throughout. 
    The second set ($a$, $\chi_{PB}^{\max}$, $\chi_{PB}^{\min}$, $\chi_{BP}$) are the unknown parameters, which we consider the variations of. 
    Their nominal values used to generate synthetic data are given.
    The final set of five values then provides the initial condition values used for simulations, unless otherwise specified.
    }
    \label{tab:Modelparameters}
\end{table}

\subsection{Code availability}

The code used throughout is available on GitHub at:\\

https://github.com/StephWilleniams/BiofilmModelAnalysis\\

Where possible, fixed random seeding was used for all synthetic data in the figures to ensure reproducibility of all the data.
Where practical, Julia was used to execute the calculations.
The ModelAnalysis module was developed during the work and provides much of the functionality used for practical identifiability and genetic algorithm development.
For convenience, some tasks required execution using MATLAB or Python.
These tasks utilised the simple vectorisation and parallelisation capabilities that MATLAB allows.

\end{document}